\documentclass[twocolumn]{jpsj2} %% two-column layout
\def\vecS{{\vec S}}

\def\gsim{\,$\raise0.3ex\hbox{$>$}\llap{\lower0.8ex\hbox{$\sim$}}$\,}
\def\lsim{\,$\raise0.3ex\hbox{$<$}\llap{\lower0.8ex\hbox{$\sim$}}$\,}
\def\Hno{({\rm H},{\rm no})}
\def\LDno{({\rm LD},{\rm no})}
\def\HLD{({\rm H},{\rm LD})}
\def\lsim{\,$\raise0.25ex\hbox{$<$}\llap{\lower0.85ex\hbox{$\sim$}}$\,}
\def\gsim{\,$\raise0.25ex\hbox{$>$}\llap{\lower0.85ex\hbox{$\sim$}}$\,}

\mathchardef\calH="0248

\title{Finite-Field Ground State of the ${{\mib S}{\mib =}{\mib 1}}$
Antiferromagnetic-Ferromagnetic Bond-Alternating Chain}

\author{Takashi \textsc{Tonegawa}$^{1}$, Kiyomi \textsc{Okamoto}$^{2}$ and
Makoto \textsc{Kaburagi}$^{3}$}

\inst{$^{1}$Department of Mechanical Engineering, Fukui University of
Technology, Fukui 910-8505, Japan \\
$^{2}$Department of Physics, Tokyo Institute of Technology, Tokyo 152-8551,
Japan \\
$^{3}$Faculty of Cross-Cultural Studies, Kobe University, Kobe 657-8501,
Japan}

\abst{We investigate the finite-field ground state of the $S\!=\!1$
antiferromagnetic-ferromagnetic bond-alternating chain described by the
Hamiltonian
${\calH}=\sum\nolimits_{\ell}\bigl\{\vecS_{2\ell-1}\cdot\vecS_{2\ell}
\!+\!J\vecS_{2\ell}\cdot\vecS_{2\ell+1}\bigr\}
\!+\!D\sum\nolimits_{\ell}\,\bigl(S_{\ell}^z)^2
\!-\!H\textstyle\sum\nolimits_\ell\,S_\ell^z$, where \hbox{$J\!\leq\!0$} and
\hbox{$-\infty\!<\!D\!<\!\infty$}.  We find that two kinds of magnetization
plateaux at a half of the saturation magnetization, the $\frac{1}{2}$-plateaux,
appear in the ground-state magnetization curve; one of them is of the Haldane
type and the other is of the large-$D$-type.  We determine the
$\frac{1}{2}$-plateau phase diagram on the $D$ versus $J$ plane, applying
the twisted-boundary-condition level spectroscopy methods developed by
Kitazawa and Nomura.  We also calculate the ground-state magnetization curves
and the magnetization phase diagrams by means of the density-matrix
renormalization-group method.}

\kword{$S\!=\!1$ antiferromagnetic-ferromagnetic bond-alternating chain,
numerical methods, ground-state magnetization curve, $\frac{1}{2}$-plateau
phase diagram, magnetization phase diagram}

\begin{document}
\maketitle

\section{Introduction} 

There has been a considerable current interest in the study of the
ground-state properties of various one-dimensional quantum spin systems in a
finite magnetic field.  In this paper we discuss the case of the $S\!=\!1$
bond-alternating chain, for which we express the Hamiltonian in the following
form:
\begin{eqnarray}
 {\calH}&&\!\!\!\!\!\!\!\!\!\!={\calH}_{\rm ex} + {\calH}_{\rm Z}\,,        \\
  && {\calH}_{\rm ex}
   = \sum\nolimits_{\ell} \Bigl\{\vecS_{2\ell-1}\cdot\vecS_{2\ell}
                  {} + J \vecS_{2\ell}\cdot\vecS_{2\ell+1}\Bigr\} \nonumber \\
  &&\qquad\qquad  {} + D \sum\nolimits_{\ell}\,\bigl(S_{\ell}^z)^2\,, \\
  && {\calH}_{\rm Z} = - H \sum\nolimits_\ell\,S_\ell^z\,,
\end{eqnarray}
\noindent
where $\vecS_\ell$ is the \hbox{$S\!=\!1$} operator at the $\ell$th site; $J$
(\hbox{$-\infty\!<\!J\!<\!\infty$}) is the parameter representing the bond
alternation of the nearest-neighbor interactions; $D$
(\hbox{$-\infty\!<\!D\!<\!\infty$}) is the uniaxial single-ion-type anisotropy
constant; $H$ (\hbox{$H\!\geq\!0$}) is the magnitude of the external
magnetic field applied along the $z$-direction.  Hereafter, we denote by $M$
the $z$ component of the total spin
$\vecS_{\rm tot}\!\equiv\!\sum_{\ell=1}^N \vecS_\ell$, where $N$, being
assumed to be a multiple of four, is the total number of spins in the system.

The finite-field as well as the zero-field ground-state properties of the
present system in the case of the antiferromagnetic-antiferromagnetic
bond-alternating chain with \hbox{$J\!>\!0$} have already been
investigated by using mainly numerical methods.~\cite{TNK,CHS}  As for the
zero-field properties, the ground-state phase diagram on the $J$ versus $D$
plane, in which the N{\'e}el, Haldane, large-$D$, and dimer phases appear,
has been determined.~\cite{TNK,CHS}  On the other hand, the results for the
ground-state magnetization curve show that the magnetization plateau at a half
of the saturation magnetization, which is called the ${1\over 2}$-plateau,
appears for arbitrary values of $J$ except for $J\!=\!1$ at least when
$D\!\geq\!0$.~\cite{TNK}

In this study we explore the finite-field ground-state properties of the
system in the case of the antiferromagnetic-ferromagnetic bond-alternating
chain with \hbox{$J\!<\!0$}, focusing our attention mainly upon the
$\frac{1}{2}$-plateau appearing in the magnetization curve.  When $D\!=\!0$,
the present system is mapped onto the isotropic \hbox{$S\!=\!2$} uniform
antiferromagnetic chain in the limit of \hbox{$J\!\to\!-\infty$}.  Thus, it
is considered that, no plateau appears in the ground-state magnetization
curve in this limit when \hbox{$D\!=\!0$}.  As we have already
discussed,~\cite{TNK} on the other hand, the ${1\over 2}$-plateau appears
when \hbox{$J\!=\!0$} and \hbox{$D\!>\!-\frac{2}{3}$}.  These imply that the
^^ ${1\over 2}$-plateau'-^^ no-plateau' transition occurs at a finite value
of $J$ for a certain region of $D$ including \hbox{$D\!=\!0$}.  Furthermore,
as has already been shown,~\cite{KO} the magnetization plateau at one third
of the saturation magnetization, the ${1\over 3}$-plateau, appearing in the
ground-state magnetization curve of the \hbox{$S\!=\!{3\over 2}$} uniform
antiferromagnetic chain with single-ion-type anisotropy changes its character
from the Haldane-type to the large-$D$-type at a finite value of the
anisotropy constant.  We expect that this type of
^^ Haldane-type-${1\over 2}$-plateau'-^^ large-$D$-type-${1\over 2}$-plateau'
transition may take place also in the present $S\!=\!1$
antiferromagnetic-ferromagnetic bond-alternating chain.  To discuss
quantitatively these transitions and to complete the ${1\over 2}$-plateau
phase diagram on the $D$ versus $J$ plane are our main purposes.  We also
aim at calculating the magnetization phase diagram as well as the ground-state
magnetization curve.

In the next section ({\S}2) we discuss the ${1\over 2}$-plateau phase
diagram.  In determining the phase diagram, we apply the
twisted-boundary-condition level spectroscopy \hbox{(TBCLS)} analyses of the
numerical diagonalization data, developed by Kitazawa~\cite{K} and also by
Nomura and Kitazawa.~\cite{NK}  Then, we discuss in {\S}3 the ground-state
magnetization curve, which we calculate by using the density-matrix
renormalization-group (DMRG) method proposed originally by
White.~\cite{White}  We also discuss the magnetization phase diagram which
is obtained from the result for the magnetization curve.  Finally, concluding
remarks are given in {\S}4.

\section{${{\mib 1}\over{\mib 2}}$-Plateau Phase Diagram}

\begin{figure}[t]
\begin{center}
\includegraphics[width=7.0cm]{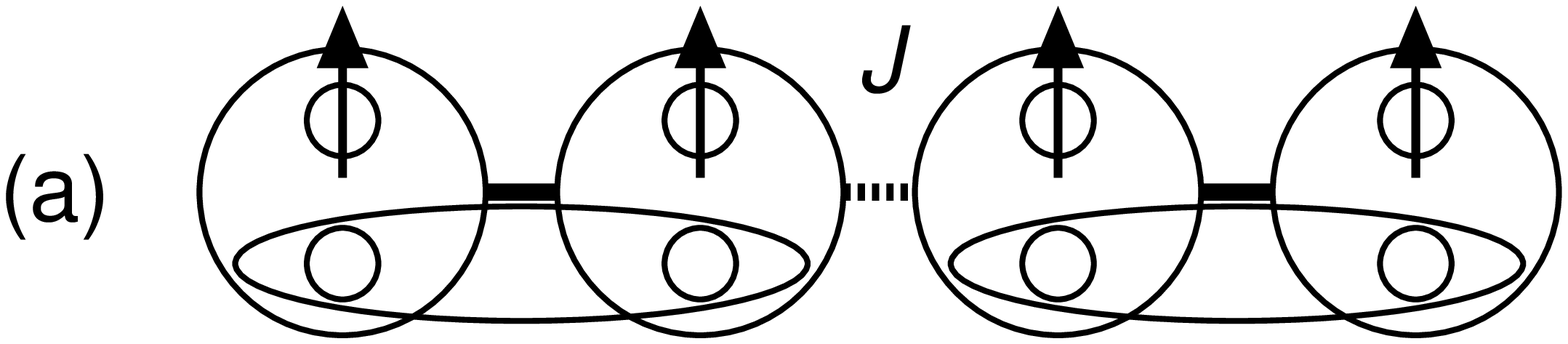}
\vspace*{0.8cm}
\end{center}
\begin{center}
\includegraphics[width=7.0cm]{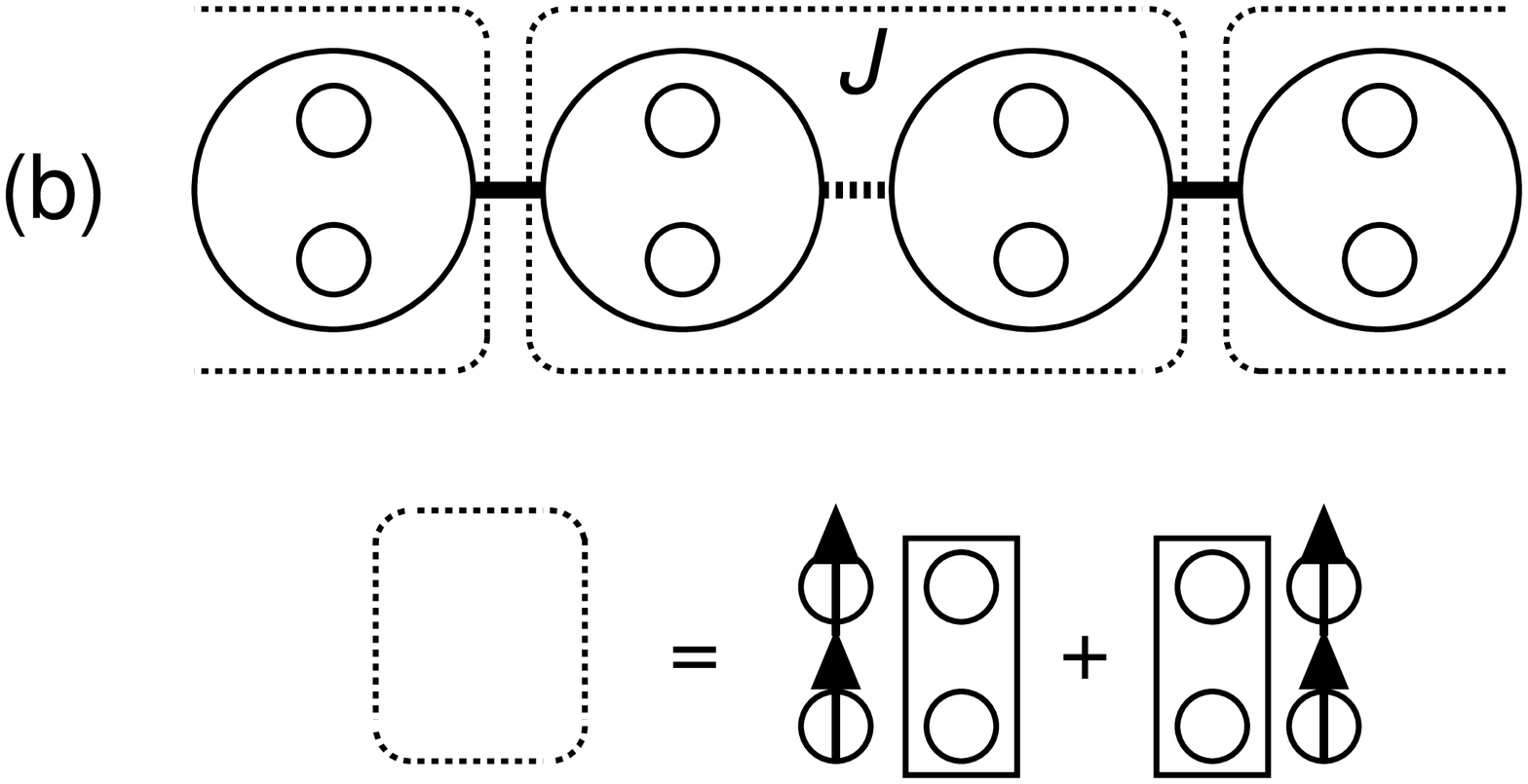}
\end{center}
\caption{Schematical pictures representing (a) the
^^ Haldane-type-${1\over 2}$-plateau' phase and (b) the
^^ large-$D$-type-${1\over 2}$-plateau' phase.  The thick solid and dotted
lines stand, respectively, for the antiferromagnetic and ferromagnetic bonds, 
and the small open circles for the \hbox{$S\!=\!\frac{1}{2}$} spins.  Each
large open circle surrounding two \hbox{$S\!=\!\frac{1}{2}$} spins stands for
an operation of constructing an \hbox{$S\!=\!1$} spin by symmetrization.  The
\hbox{$S\!=\!\frac{1}{2}$} spin with an upward arrow is in the up state
$\uparrow$.  Furthermore, two \hbox{$S\!=\!\frac{1}{2}$} spins in a flat
ellipse form a singlet dimer
\hbox{$\frac{1}{\sqrt{2}}
\bigl(\uparrow\downarrow\!-\!\downarrow\uparrow\bigr)$}, and
two \hbox{$S\!=\!\frac{1}{2}$} spins in a rectangular form a triplet dimer
\hbox{$\frac{1}{\sqrt{2}}
\bigl(\uparrow\downarrow\!+\!\downarrow\uparrow\bigr)$.}}
\label{fig:1}
\end{figure}

From the considerations presented in the previous section, it is expected that
three phases, i.e., the ^^ no-plateau',
^^ Haldane-type-${1\over 2}$-plateau', and
^^ large-$D$-type-${1\over 2}$-plateau' phases appear in the
${1\over 2}$-plateau phase diagram.  Schematical pictures in terms of a kind
of the valence-bond-solid picture, which represents the latter two phases, are
shown in Fig.~\ref{fig:1}.

We can determine the transition lines between two of the above thee phases
rather precisely by applying the TBCLS methods in the following
way.  Both the ^^ Haldane-type-${1\over 2}$-plateau'-^^ no-plateau' transition
and the ^^ large-$D$-type-${1\over 2}$-plateau'-^^ no-plateau' transition are
the Berezinskii-Kosterlitz-Thouless transition~\cite{BKT} accompanying no
spontaneous breaking of the translational symmetry of the Hamiltonian
$\calH$.  Therefore, the critical points $J_{\rm c}^{\Hno}$ and
$J_{\rm c}^{\LDno}$ of $J$, which are, respectively, for the former and
latter transitions, for a given value of $D$ (or, $D_{\rm c}^{\Hno}$ and
$D_{\rm c}^{\LDno}$ of $D$ for a given value of $J$) can be estimated by means
of Nomura and Kitazawa's TBCLS method.~\cite{NK}  We denote by
$E_0^{({\rm P})}(N,M)$ the lowest energy eigenvalue of $\calH_{\rm ex}$ with
the periodic boundary condition within the subspace determined by the values
of $N$ and $M$, and also by $E_0^{({\rm T})}(N,M,P)$ the lowest energy
eigenvalue of $\calH_{\rm ex}$ with the twisted boundary condition within the
subspace determined by the values of $N$, $M$, and $P$, where $P$ is the
eigenvalue of the space inversion operator.  Then, Nomura and Kitazawa's
TBCLS method~\cite{NK} predicts that $J_{\rm c}^{\Hno}$ ($D_{\rm c}^{\Hno}$)
is obtained by extrapolating
$J_{\rm c}^{\Hno}(N)$ $\bigl(D_{\rm c}^{\Hno}(N)\bigr)$ determined from the
equation
\begin{eqnarray}
  E_0^{({\rm T})}&&\!\!\!\!\!\!\!\!\!\!\!\!\!\!\Bigl(N,\frac{N}{2},-1\Bigr)
                                                              \nonumber  \\
     &&\!\!\!\!\!\!\!\!\!\!=\frac{1}{2}
       \Bigl[E_0^{({\rm P})}\Bigl(N,\frac{N}{2}+2\Bigr)
           + E_0^{({\rm P})}\Bigl(N,\frac{N}{2}-2\Bigr)\Bigr]
\end{eqnarray}
to \hbox{$N\!\to\!\infty$}.  Similarly,
$J_{\rm c}^{\LDno}$ ($D_{\rm c}^{\LDno}$) is obtained by extrapolating
$J_{\rm c}^{\LDno}(N)$ $\bigl(D_{\rm c}^{\LDno}(N)\bigr)$ determined from
\begin{eqnarray}
  E_0^{({\rm T})}&&\!\!\!\!\!\!\!\!\!\!\!\!\!\!\Bigl(N,\frac{N}{2},+1\Bigr)
                                                              \nonumber  \\
     &&\!\!\!\!\!\!\!\!\!\!=\frac{1}{2}
       \Bigl[E_0^{({\rm P})}\Bigl(N,\frac{N}{2}+2\Bigr)
           + E_0^{({\rm P})}\Bigl(N,\frac{N}{2}-2\Bigr)\Bigr]
\end{eqnarray}
to \hbox{$N\!\to\!\infty$}.

On the other hand, the
^^ Haldane-type-${1\over 2}$-plateau'-^^ large-$D$-type-${1\over 2}$-plateau'
transition is expected to be of the Gaussian type as in the case of the
^^ Haldane-type-${1\over 3}$-plateau'-^^ large-$D$-type-${1\over 3}$-plateau'
transition in the the \hbox{$S\!=\!{3\over 2}$} uniform antiferromagnetic chain
with single-ion-type anisotropy,~\cite{KO} discussed in the previous
section.  Thus, we can apply Kitazawa's TBCLS method~\cite{K} to estimate the
critical point $J_{\rm c}^{\HLD}$ of $J$ for a given $D$
(or, $D_{\rm c}^{\HLD}$ for a given $J$) in this transition.  According to
this method, $J_{\rm c}^{\HLD}$ ($D_{\rm c}^{\HLD}$) can be estimated by
extrapolating $J_{\rm c}^{\HLD}(N)$ $\bigl(D_{\rm c}^{\HLD}(N)\bigr)$
determined from
\begin{equation}
      E_0^{({\rm T})}\Bigl(N,\frac{N}{2},+1\Bigr)
              = E_0^{({\rm T})}\Bigl(N,\frac{N}{2},-1\Bigr)
\end{equation}
to \hbox{$N\!\to\!\infty$}.  It is noted that in the limit of
\hbox{$N\!\to\!\infty$}, the value of
$E_0^{({\rm T})}\Bigl(N,\frac{N}{2},-1\Bigr)$ is smaller or larger than that
of $E_0^{({\rm T})}\Bigl(N,\frac{N}{2},+1\Bigr)$ depending upon whether in
the ^^ Haldane-type-${1\over 2}$-plateau' region or in the
^^ large-$D$-type-${1\over 2}$-plateau' region.

\begin{figure}[t]
\begin{center}
\includegraphics[width=7.0cm]{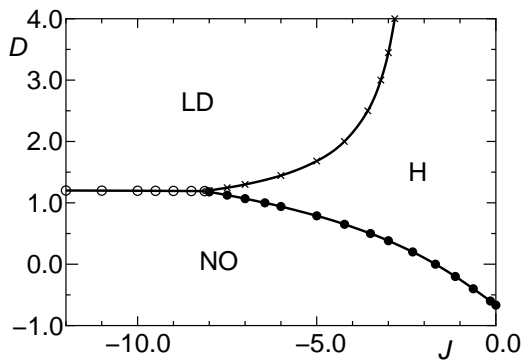}
\end{center}
\caption{${1\over 2}$-plateau phase diagram on the $D$ versus $J$
plane.  Here, NO, H, and LD stand, respectively, for the ^^ no-plateau',
^^ Haldane-type-${1\over 2}$-plateau', and
^^ large-$D$-type-${1\over 2}$-plateau' regions.  The three transition
lines meet at the point (\hbox{$J\!=\!-8.13\pm0.01$},
\hbox{$\,$$D\!=\!1.19\pm0.01$}), which is marked by the double circle.}
\label{fig:2}
\end{figure}

Determining the ${1\over 2}$-plateau phase diagram on the $D$ versus $J$
plane, we have first calculated numerically the energy eigenvalues
\hbox{$E_0^{({\rm P})}\Bigl(N,\frac{N}{2}\!+\!2\Bigr)$},
\hbox{$E_0^{({\rm P})}\Bigl(N,\frac{N}{2}\!-2\!\Bigr)$},
$E_0^{({\rm T})}\Bigl(N,\frac{N}{2},+1\Bigr)$, and
$E_0^{({\rm T})}\Bigl(N,\frac{N}{2},-1\Bigr)$ for a variety values of $J$ and
$D$ for finite-size systems with \hbox{$N\!=\!8$}, $12$, and $16$ spins,
employing the computer program package KOBEPACK~\cite{KP} coded by means of
the Lancz{\"o}s technique.  Then, we have solved numerically eqs.$\,$(4),
(5), and (6), respectively, to evaluate $J_{\rm c}^{\Hno}(N)$
$\bigl(D_{\rm c}^{\Hno}(N)\bigr)$,
$J_{\rm c}^{\LDno}(N)$ $\bigl(D_{\rm c}^{\LDno}(N)\bigr)$, and
$J_{\rm c}^{\HLD}(N)$ $\bigl(D_{\rm c}^{\HLD}(N)\bigr)$.  Finally, we have
extrapolated these finite-size values to \hbox{$N\!\to\!\infty$} to estimate
$J_{\rm c}^{\Hno}$ ($D_{\rm c}^{\Hno}$), $J_{\rm c}^{\LDno}$ 
($D_{\rm c}^{\LDno}$), and $J_{\rm c}^{\HLD}$ ($D_{\rm c}^{\HLD}$) by
assuming that the $N$-dependences of the finite-size values are quadratic
functions of $N^{-2}$.  Plotting the extrapolated results on the $D$ versus
$J$ plane, we have obtained the $\frac{1}{2}$-plateau phase diagram as
depicted in Fig.~\ref{fig:2}.  It is noted that, for example,
\hbox{$J_{\rm c}^{\Hno}\!=\!-1.69\pm0.01$} for $D\!=\!0$.  Furthermore, the
three transition lines in the phase diagram meet at the point
($J\!=\!-8.13\pm0.01$,$\,$$D\!=\!1.19\pm0.01$).

As can be seen from Fig.~\ref{fig:2}, the
^^ large-$D$-type-${1\over 2}$-plateau'-^^ no-plateau' transition line
is almost parallel to the $J$-axis in the whole region of
$J\!\lsim\!-8.13$.  This result may be attributed to the fact that the
present \hbox{$S\!=\!1$} system in this region is almost equivalent to the
\hbox{$S\!=\!2$} uniform antiferromagnetic chain with single-ion-type
anisotropy, since $\vert J\vert$ is fairly large with \hbox{$J\!<\!0$}.  Thus,
we may expect a direct transition from the ^^ no-plateau' phase to the
^^ large-$D$-type-${1\over 2}$-plateau' one in the above \hbox{$S\!=\!2$}
chain, in contrast to the case of the ${1\over 3}$-plateau in the corresponding
$S\!=\!{3\over 2}$ chain,~\cite{KO} mentioned in {\S}1.  Figure 2 also shows
that the ^^ large-$D$-type-${1\over 2}$-plateau' phase appears when both
$\vert J\vert$ ($J\!<\!0$) and $D$ ($D\!>\!0$) are sufficiently large.  The
fitting of several values of $J_{\rm c}^{\HLD}$ for sufficiently large $D$'s
to a quadratic function of $D^{-1}$ suggests that the
^^ Haldane-type-${1\over 2}$-plateau'-^^ large-$D$-type-${1\over 2}$-plateau'
transition line approaches $J\!=\!-2.05\pm0.05$ in the limit of
$D\!\to\!\infty$.

%$\lim_{D\to\infty}J_{\rm c}^{\HLD}\!=\!-0.21\pm0.1$.

\section{Magnetization Curve and Magnetization Phase Diagram}

Let us begin with discussing the saturation field $H_{\rm s}$ for the present
system in the case of $J\!<\!0$.  It is given by
\begin{equation}
    H_{\rm s} = E_0^{({\rm P})}(N,N) - E_0^{({\rm P})}(N,N-1)
              = D + 2
\end{equation}
or
\begin{equation}
  H_{\rm s}
    = \frac{1}{2} \lim_{N\to\infty}\Big\{E_0^{({\rm P})}(N,N) 
                                       - E_0^{({\rm P})}(N,N-2)\Bigr\}
    = \frac{1}{2} E_{\rm B}
\end{equation}
depending on whether $2(D\!+\!2)\!>\!E_{\rm B}$ or $2(D\!+\!2)\!<\!E_{\rm B}$,
where $E_{\rm B}$ is the maximum energy of the ferromagnetic-two-magnon bound
state.  In principle, the energy $E_{\rm B}$ can be calculated
analytically.  However, this calculation in the case of $J\!\ne\!1$ has not
yet been done as far as we know,~\cite{T} and therefore we estimate
$H_{\rm s}$ numerically when $2(D\!+\!2)\!<\!E_{\rm B}$.

We have applied the DMRG method~\cite{White} to calculate the ground-state
magnetization curve, which we define here as the average magnetization
\hbox{$m(N)(\equiv\!M/N)$} per spin versus the reduced field $H/H_{\rm s}$
curve.  In this application, we usually have to impose open boundary
conditions for the Hamiltonian treated.  In the present system, we have two
kinds of open boundary conditions, Oa and Of, which are, respectively,
the cases where the edge bonds are the antiferromagnetic and ferromagnetic
bonds.  We adopt, respectively, the Oa and Of conditions when we calculate the
magnetization curves in the ^^ Haldane-type-${1\over 2}$-plateau' and
^^ large-$D$-type-${1\over 2}$-plateau' regions.  This is because, as can
be seen from the schematical pictures for these phases shown in
Fig.~\ref{fig:1}, no degrees of freedom appear at both edges of the chain in
the above adoption, which leads to a less boundary effect.

We have calculated the lowest energy eigenvalue $E_0^{({\rm Ox})}(N,M)$
(x$=$f or a) under the open boundary condition Ox of $\calH_{\rm ex}$
within the subspace determined by the values of $N$ and $M$ for
\hbox{$N\!=\!36$}, $64$, and $84$ and for all values of $M$ (\hbox{$M\!=\!0$},
$1$, $\cdots$, $N$).  From these results the ground-state magnetization curves
for \hbox{$N\!=\!36$}, $64$, and $84$ can be readily obtained.~\cite{BF}   The
resulting magnetization curve for each $N$ is a stepwisely increasing function
of $H$, starting from \hbox{$m(N)\!=\!0$} at \hbox{$H\!=\!H_0(N)$} and
reaching to the saturation value \hbox{$m(N)\!=\!1$} at
\hbox{$H\!=\!H_{\rm s}(N)$}.  We note that
$H_{\rm s}\!=\!\lim_{N\to\infty}H_{\rm s}(N)$.

\begin{figure}[t]
\begin{center}
\includegraphics[width=7.0cm]{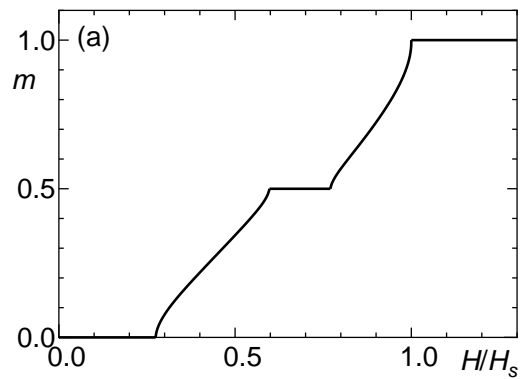}
\end{center}
\vspace*{0.5cm}
\begin{center}
\includegraphics[width=7.0cm]{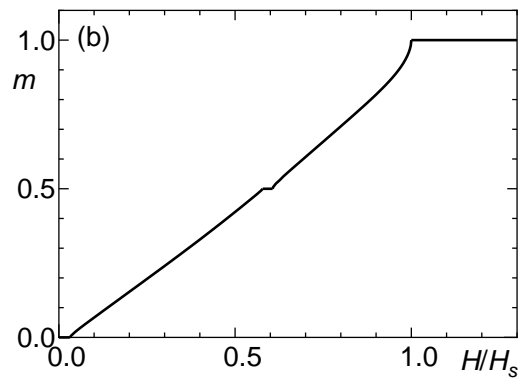}
\end{center}
\caption{Ground-state magnetization curves in the \hbox{$N\!\to\!\infty$}
limit obtained for the cases of (a) \hbox{$J\!=\!-1.0$} and \hbox{$D\!=\!3.0$}
and of (b) \hbox{$J\!=\!-10.0$} and \hbox{$D\!=\!3.0$}}
\label{fig:3}
\end{figure}

Except for plateau regions, a satisfactorily good approximation to the
magnetization curve in the \hbox{$N\!\to\!\infty$} limit, i.e., the
\hbox{$m\bigl(\equiv\!m(\infty)\bigr)$} versus $H/H_{\rm s}$ curve, may be
obtained by drawing a smooth curve through the midpoints of the steps in the
finite-size magnetization curves.~\cite{BF}  As for the estimations of
\hbox{$H_0\bigl(\equiv\!H_0(\infty)\bigr)$}, which is nothing but the energy
gap between the ground state and a first excited state in the case of
\hbox{$H\!=\!0$}, as well as of $H_1$ and $H_2$, which are the lowest and
highest values of $H$ giving the $\frac{1}{2}$-plateau, respectively, we
extrapolate the corresponding results for \hbox{$N\!=\!36$}, $64$, and $84$
to \hbox{$N\!\to\!\infty$}, assuming again the $N$-dependences of the
finite-size values are quadratic functions of $N^{-2}$.  The magnetization
curve thus obtained for the cases of \hbox{$J\!=\!-1.0$} and
\hbox{$D\!=\!3.0$} and of \hbox{$J\!=\!-10.0$} and \hbox{$D\!=\!3.0$},
which are, respectively, in the ^^ Haldane-type-${1\over 2}$-plateau'
and ^^ large-$D$-type-${1\over 2}$-plateau' regions (see Fig.~\ref{fig:2}),
are depicted in Fig.~\ref{fig:3}.

\begin{figure}[t]
\begin{center}
\includegraphics[width=7.0cm]{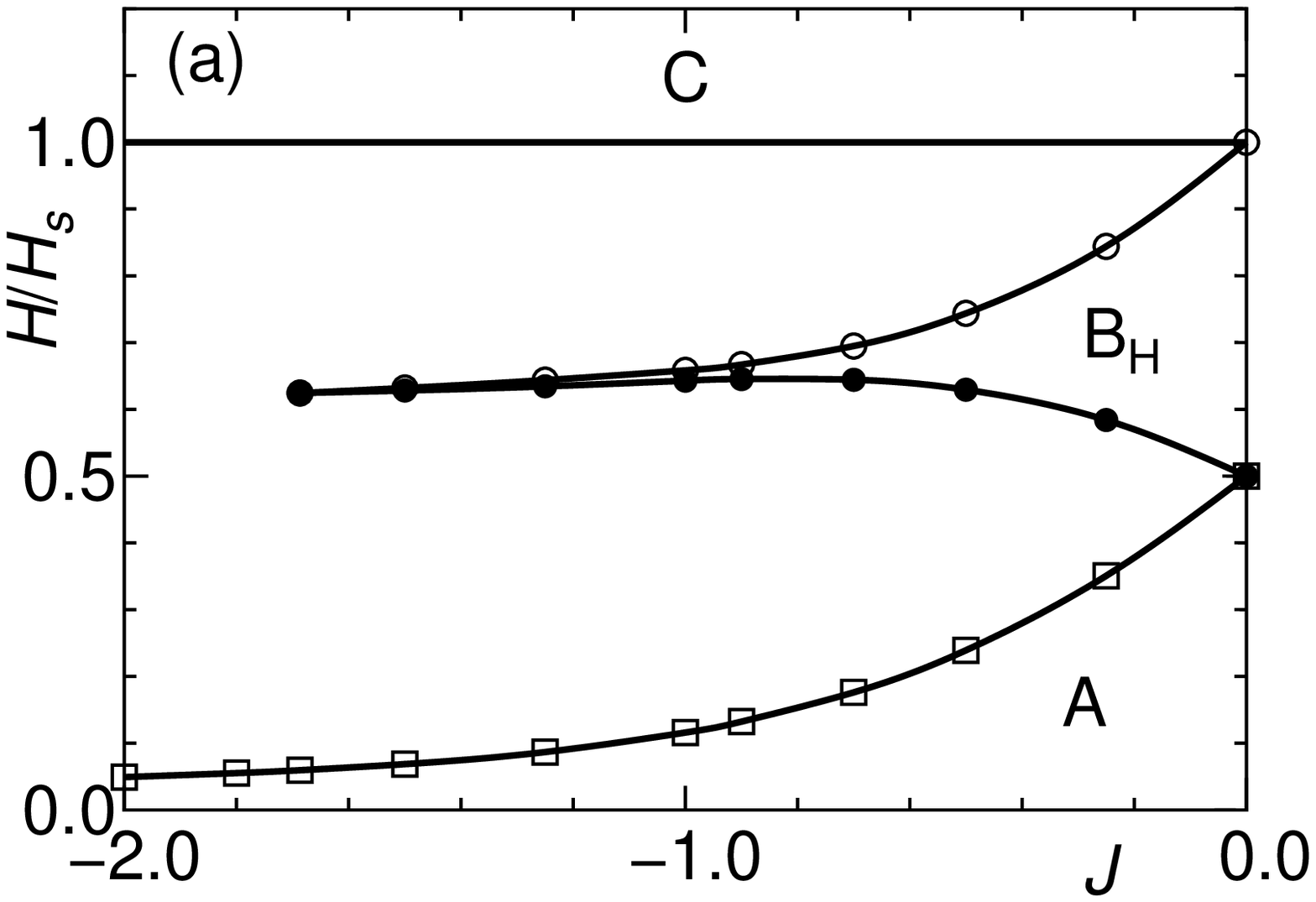}
\end{center}
\vspace{0.5cm}
\begin{center}
\includegraphics[width=7.0cm]{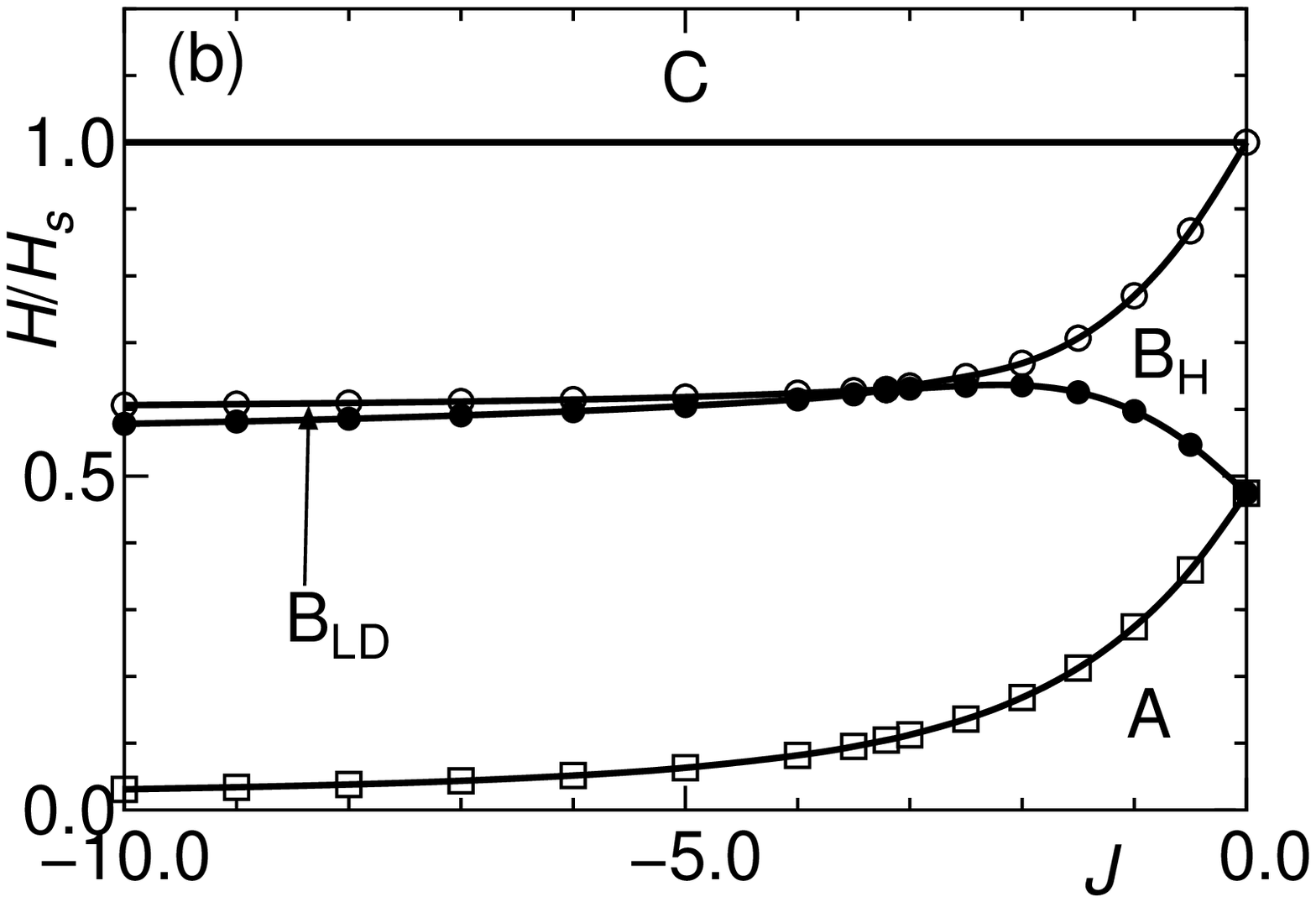}
\end{center}
\caption{Magnetization phase diagrams on the $H/H_{\rm s}$ versus $J$
plane obtained for the cases of (a) \hbox{$D\!=\!0.0$} and (b)
\hbox{$D\!=\!3.0$}, where $H_0/H$ (open squares), $H_1/H$ (solid
circles), and $H_2/H$ (open circles) ate plotted as functions of
$J$.  In the regions A, B$_{\rm X}$ (X$=$H or LD), and C, the average
magnetization $m$ per spin is equal to $0$ (the $0$-plateau region),
$\frac{1}{2}$ (the $\frac{1}{2}$-plateau region), and $1$ (the saturated
magnetization region); the B$_{\rm H}$ and B$_{\rm LD}$ regions are,
respectively, the ^^ Haldane-type-${1\over 2}$-plateau' and
^^ large-$D$-type-${1\over 2}$-plateau' regions.  In other regions,
$m$ increases continuously with the increase of $H/H_{\rm s}$.}
\label{fig:4}
\end{figure}

The magnetization phase diagram on the $H/H_{\rm s}$ versus $J$ plane for
a given value of $D$ is obtained by plotting $H_0$, $H_1$, $H_2$, and
$H_{{\rm s}}$ as a function of $J$.  The results for the cases of
\hbox{$D\!=\!0.0$} and \hbox{$D\!=\!3.0$} are shown in
Fig.~\ref{fig:4}.  Figure~\ref{fig:4}(a) demonstrates that in the case of
\hbox{$D\!=\!0.0$}, the $\frac{1}{2}$-plateau,
which is of the ^^ Haldane-type', appears when 
\hbox{$-1.69\pm0.01<\!J\!\leq0.0$}, and Fig.~\ref{fig:4}(b) shows that in
the case of \hbox{$D\!=\!3.0$}, the $\frac{1}{2}$-plateau disappears at the
critical point \hbox{$J_{\rm c}^{\HLD}\!=\!-3.21\pm0.01$} for the
^^ Haldane-type-${1\over 2}$-plateau'-^^ large-$D$-type-${1\over 2}$-plateau'
phase transition (see Fig.~\ref{fig:2}).

\section{Concluding Remarks}

We have investigated the finite-field ground state of the $S\!=\!1$
antiferromagnetic-ferromagnetic bond-alternating chain described by the
Hamiltonian ${\calH}$ [see eqs.$\,$(1-3)] with \hbox{$J\!\leq\!0$} and
\hbox{$-\infty\!<\!D\!<\!\infty$}, focusing our attention mainly upon the
$\frac{1}{2}$-plateau appearing in the magnetization curve.  We have found
that two kinds of magnetization plateaux, i.e., the
^^ Haldane-type-${1\over 2}$-plateau' and the
^^ large-$D$-type-${1\over 2}$-plateau' [Fig.~\ref{fig:1}] appear.  We have
determined the $\frac{1}{2}$-plateau phase diagram on the $D$ versus $J$
plane [Fig.~\ref{fig:2}], applying the TBCLS methods.~\cite{K,NK}   We have
also calculated, by means of the DMRG method~\cite{White}, the ground-state
magnetization curves for a variety values of $J$ with $D$ fixed at $D\!=\!0.0$
and $3.0$, two examples of which being presented [Fig.~\ref{fig:3}].  Using the
results for the magnetization curves, we finally obtained the magnetization
phase diagrams on the $H/H_{\rm s}$ versus $J$ plane for \hbox{$D\!=\!0.0$}
and $3.0$ [Fig.~\ref{fig:4}].

For the purpose of exploring the effect of frustration, we are now studying
the case where the antiferromagnetic next-nearest-neighbor interaction term
$J_2\sum_\ell\vecS_{\ell}\cdot\vecS_{\ell+2}$ (\hbox{$J_2\!>\!0$}) is added
to the Hamiltonian of eqs.$\,$(1-3), assuming either $J\!>\!0$ or
$J\!<\!0$.  It is noted that when $J\!=\!0$, the system is reduced to the
$S\!=\!1$ antiferromagnetic ladder, where the ratio of the leg interaction
constant to the rung one is $J_2$.  The results will be published in the near
future.

\section*{Acknowledgments}
We wish to thank Professor T.\ Hikihara by whom the DMRG program used in this
study is coded.  We also thank the Supercomputer Center, Institute for Solid
State Physics, University of Tokyo, the Information Synergy Center, Tohoku
University, and the Computer Room, Yukawa Institute for Theoretical Physics,
Kyoto University for computational facilities.  The present work has been
supported in part by Grants-in-Aid for Scientific Research (C)
(No.\ 16540332, No.\ 14540329, and No.\ 14540358) and a Grant-in-Aid for
Scientific Research on Priority Areas (B) (^^ Field-Induced New Quantum
Phenomena in Magnetic Systems') from the Ministry of Education, Culture,
Sports, Science and Technology.

\end{document}